\documentstyle [aps,preprint]{revtex}
\begin{document}
\title{Non-equilibrium thermodynamics of thermionic emission
 processes in abrupt semiconductor junctions, including the effects
 of surface states}

\author{G. Gomila, A.P\'erez-Madrid, J.M. Rub\'{\i}
\\Departament de F\'{\i}sica Fonamental\\Facultat de F\'{\i}sica\\
Universitat de Barcelona\\Diagonal 647, 08028 Barcelona, Spain}
\date{\today}
\maketitle
\begin{abstract}
The methods of non-equilibrium thermodynamics of systems with an 
interface
have been applied to the study of thermionic emission processes 
in abrupt semiconductor junctions, including the effects of 
 surface states . Our analysis covers different
situations of interest, concerning unipolar and 
bipolar 
systems, either 
degenerate or not, with or
without interface states and under steady and non-steady conditions.
 In this way, a complete phenomenological modelling
of thermionic emission-interface states processes has been proposed, 
overcoming some of the inherent limititations of the 
currently existing models.  
\end{abstract}
\pagebreak
\section{Introduction}

Metal-semiconductor contacts have been 
extensively studied in the literature~\cite{Sze}, ~\cite{Roderick}.
At the beginning,  basically two different approaches
were followed to describe their rectifying properties: the diffusion 
approach
~\cite{Schottky}, in which the main current limiting mechanism was the
 diffusion
of carriers through the semiconductor; and the thermionic emission 
(TE) approach
~\cite{Bethe}, in which the dominant process was the thermal 
emission of
 carriers over the contact barrier. These two approaches were 
 synthesized
later ~\cite{Sze2}, giving rise to the so-called thermionic 
emission-
diffusion (TE-D) models, in which both current limiting processes were 
taken into 
account. A similar development occurred in the more recent area of
 heterojunctions, in which both TE~\cite{Anderson}, and 
 TE-D~\cite{llibre} models were also proposed. 
 
 These models have been further 
 generalised
 by including the effects of interface  states that
 very often are present at semiconductor junctions, see for instance,
 ~\cite{Bardeen}, ~\cite{Wu}, 
 ~\cite{Milnes}. One may call these generalizations
 thermionic emission-interface state-diffusion (TE-IS-D) models.
 In some cases, mostly in the context of metal-semiconductor (MS) 
 junctions,
 a thin insulating layer was introduced to account for some non-ideal 
 properties~\cite{Sze}, ~\cite{Roderick}. Its main effect was to allow
 the electric potential to be discontinuous across the junction. 
 Therefore, in this
 paper we will distinguish between {\em abrupt} junctions 
 (continuous electric
 potential) and {\em non-abrupt} ones (discontinuous electric 
 potential).
 For many purposes, the models mentioned above suffice to describe 
 most of the properties
 of semiconductor junctions~\cite{Sze2}, ~\cite{Milnes}.
  
 Despite their wide application, a general phenomenological derivation
 of those models is still lacking in the current literature. This may 
 be due to the fact
 that the TE-IS and diffusion processes are, a priori, of a different 
 nature,
 which suggests applying to the former a microscopic-kinetic approach,
 while to the latter a phenomenological one. As a 
 consequence, this fact has caused a situation where the models
 are not formulated in a general enough way, 
 in the sense that some situations or processes 
 have not been explored, mainly those with respect to 
 the TE-IS processes.
 
 It is our purpose in this paper to present a complete 
 analysis
 of the TE-IS processes in abrupt semiconductor junctions. This
 study overcomes some of
 the inherent limitations of the previous models and provides a
 proper framework to 
 study these junctions in a more complete way.
 
 We would like to draw special attention to the fact that
 we will develop the whole analysis in the same
 phenomenological framework as where diffusion processes may be 
 treated,
 ~\cite{Noltros}, allowing us then to present a unified
 derivation of the TE-IS-D models in abrupt semiconductor 
 junctions. In this
 way all the situations and processes may be studied, including the 
 effects
 of degenerancy, bipolarity, non-stationarity or interface state 
 filling,
 which are treated only in an approximate way in most of the models
 mentioned above.
 
 The organisation of the paper is as follows: In section II, we 
 summarize
 the main results concerning non-equilibrium thermodynamic 
 description
 of semiconductor junctions. In this framework, and with the help of 
 the
 formalism of the internal degrees of freedom, we derive in 
 section III
 the TE law for abrupt unipolar junctions without interface 
 states. This
 result is further generalised in section IV, to include the effects 
 of 
 interface states or bipolarity. Finally, in section V we summarize 
 our main results.
   
\section{Non-equilibrium thermodynamic description of 
          semiconductor junctions}

When two different semiconductors (or, in general, two different 
materials)
are in contact, a very narrow region exists in between, which 
exhibits peculiar macroscopic behaviour. It has been 
shown~\cite{Mazur}, ~\cite{Bedeaux}, that
the properties of this region can be explained
in a simple way, by substituting it by a two-dimensional 
interface, with
its own macroscopic properties, and by assuming that the bulks to 
 will behave as in
the absence of the junction just up to this interface. This 
introduces a discontinuous description of the system, in which the 
relevant
fields are the extrapoled bulk fields (discontinuous through the 
interface)
and the surface fields (to describe surface 
properties),~\cite{Noltros} 
~\cite{Bedeaux}. The discontinuity in the bulk fields 
makes it necessary to introduce
some discontinuity relationships in the description of these 
systems, and
the presence of the two-dimensional interface the proper surface 
equations to describe its properties.
A systematic and thermodynamically consistent way to derive a 
complete
set of equations (bulk, interface and discontinuity equations) to 
describe 
these systems, is provided by the theory of non-equilibrium 
thermodynamics
of systems with interfaces~\cite{Mazur}, ~\cite{Degroot}. Precisely, 
by
following this approach, a one dimensional phenomenological model
for drift-diffusion processes in abrupt semiconductor junctions 
was derived in~\cite{Noltros}, which included, apart from the usual
drift-diffusion bulk equations, the corresponding surface and 
discontinuity relations.

Of particular interest for the present paper are the
discontinuity  relationships for the drift-diffusion processes, 
for they describe the TE-IS processes that take place at the 
semiconductor junction. To arrive at these discontinuity relations,
we first evaluated in ~\cite{Noltros} the surface entropy production
 corresponding
to a semiconductor junction when only drift and diffusion processes 
are considered. For a one-dimensional system one
obtains~\cite{Noltros},
\begin{equation}
\sigma^{s} = - \frac{1}{T} \vec{J}^{+} \cdot \vec{X}^{+}
             - \frac{1}{T} \vec{J}^{-} \cdot \vec{X}^{-}
             \label{intersi}
\end{equation}
for the case in which interface states are present, and
\begin{equation}
\sigma^{s} = - \frac{1}{T} \vec{J}^{+} \cdot [\vec{\phi}]_{-}
\label{interno}
\end{equation}
when they are not.
In Eqs.~(\ref{intersi}) and~(\ref{interno}), we have defined 
$\vec{J}^{\pm} = ( J_{1}^{\pm}, J_{2}^{\pm} )$, where $J_{k}^{\pm}$ 
stands
for the perpendicular component to the interface of the current 
density
number of electrons (k=1) and holes (k=2), evaluated at the left ($-$)
 and
right ($+$) hand sides of the junction. Similarly, we have  
$\vec{X}^{\pm} = ( X_{1}^{\pm}, X_{2}^{\pm})$ ,  $[\vec{\phi}]_{-} =
([\phi_{1}]_{-}, [\phi_{2}]_{-})$, with $X_{k}^{+} = 
\phi_{k}^{+}-\phi_{k}^{s}$, $X_{k}^{-} = \phi_{k}^{s}-\phi_{k}^{-}$, 
and
$[\phi_{k}]_{-} = \phi_{k}^{+}-\phi_{k}^{-}$, with
$\phi_{k}^{\pm, s}$ being
the corresponding electrochemical potentials (quasi-Fermi levels) 
for the k-th carrier in the bulks ($\pm$),
or at the interface (s), defined as
\begin{equation}
\phi_{k}^{\pm} = \mu_{k}^{\pm} + q_{k} V^{\pm},  \hspace{1 cm}
\phi_{k}^{s} = \mu_{k}^{\pm} + q_{k} [V]_{+}; \hspace{0.5 cm} k=1,2  
\end{equation}
where $\mu_{k}^{\pm, s}$ are the chemical potentials (Fermi levels), 
$q_{k}$ the charge of the k-th carrier, and $V^{\pm}$ the electric 
potential
with $[V]_{+} = \frac{1}{2} (V^{+} + V^{-})$. All the magnitudes 
defined
above are evaluated at the interface, that is, at $x=0$.
For abrupt junctions, the electric potential is continuous across 
the junction,
 then one has $[V]_{-} = 0$.

>From these entropy productions the following discontinuty relations
may be derived~\cite{Noltros},
\begin{equation}
\vec{J}^{\pm} = -\frac{1}{T} \vec{F}^{\pm} (\vec{X}^{+}, \vec{X}^{-})
 \label{F+-}
\end{equation}
which is valid when interface states are present. Otherwise, the 
corresponding
relation is
\begin{equation}
 \vec{J}^{+} = -\frac{1}{T} \vec{F} ( [\vec{\phi}]_{-})
 \label{F}
\end{equation}
In the above equations, $\vec{F}^{\pm}$ and $\vec{F}$ are 
some unspecified functions 
satisfying the equilibrium condition
\begin{equation}
\vec{F}^{\pm}( \vec{0}, \vec{0} )= \vec{F} (\vec{0} ) = 0 
  \label{equil} 
\end{equation}
the second principle of thermodynamics
\begin{eqnarray}
& & \vec{X}^{+} \cdot \vec{F}^{+} + \vec{X}^{-} \cdot 
 \vec{F}^{-} \ge 0  
\label{nonequil1} \\
& & [\vec{\phi}]_{-} \cdot \vec{F} \ge 0
\label{nonequil2}
\end{eqnarray}
and the Onsager's symmetry relations
\begin{equation}
\left. \frac{\partial F_{k}^{\alpha}}
         {\partial X_{j}^{\beta}} \right|_{eq} =
\left. \frac{\partial F_{j}^{\beta}}
         {\partial X_{k}^{\alpha}}\right|_{eq}
\hspace{0.4 cm}  {k, \, j = 1, 2 \atop 
\alpha, \, \beta = +, -}
\label{Onsager1}
\end{equation}

\begin{equation}
\left. \frac{\partial F_{1}}{\partial [\phi_{2}]_{-}} \right|_{eq} = 
\left. \frac{\partial F_{2}}{\partial [\phi_{1}]_{-}} \right|_{eq}   
\hspace{1 cm} \mbox{.} \label{Onsager2}                            
\end{equation}
Finally, it should be noted that the functions $\vec{F}^{\pm}$  
($\vec{F}$) may also
depend on the bulk number densities, $n_{k}^{\pm}$  
 (surface number density $n_{k}^{s}$) and on temperature,
 T ~\cite{Noltros}. 

For the case under consideration, in which recombination-generation
or tunnelling  processes are neglected, one usually assumes that 
the transport of
current across the interface is controlled by 
the thermal excitation of
 carriers over the barrier (or barriers) that appears at the 
junction. 
For this reason, we will call Eqs.~(\ref{F+-}) 
and~(\ref{F}) the TE-IS and TE 
relations, respectively and the search for explicit
expressions for them is the main purpose of next sections.

\section{ Thermionic emission relation for unipolar abrupt junctions}

As commented previously, the 
derivation of all these expressions will be carried out by
 means of the methods of
non-equilibrium thermodynamics. Due to the intrinsic non-linear nature
of the emission processes,~\cite{Noltros}, one has to go beyond the 
usual
linear phenomenological relations, and propose more general ones. 
One way to systematically derive such a non-linear relations, still 
under
 the framework of non-equilibrium thermodynamics, is provided by the 
 formalism
 of the internal degrees of freedom. 
 This will be the approach we will follow in this paper. For the sake
 of clarity, we will devote the next subsection to introduce the 
 internal
degrees of freedom in the context of semiconductor junctions, 
postponing
the derivation of the TE relation for next subsection.

\subsection{Internal degrees of freedom and semiconductor junctions}

In the thermodynamic description of a system one has to distinguish
between transport and relaxation processes. For the former,
 transport equations accounting for the evolution
of  quantities,as density, velocity, etc. can be formulated 
starting from the conservation laws of mass, momentum, etc.
Transport equations then describe the evolution of those
quantities, essentialy due to two mechanisms namely, diffusion
and convection. Relaxation processes involve transformations
ocurring at each point of the system whose mechanism is 
intrinsically non-linear and is usually modelled by the pass through
a potential barrier. In non-equilibrium thermodynamics language
this process may be viewed as diffusion in an internal space,
the space of the internal variable. These scheme has been 
satisfactorily applied to a number of physical 
situations~\cite{Reaccions}, ~\cite{Ignasi},~\cite{MovBrounia}
~\cite{Degroot}.

Concerning semiconductor junctions the following example will
illustrate the way of introducing internal variables to arrive
at a complete characterization of the system.
Let us consider the emission of a 
carrier
through the interface. When the carrier reaches the interface, it has 
several options: to jump directly into the other side of the junction,
or into an empty surface state, or simply to go back. The net effect 
of these
processes is known to be  non-linear~\cite{Noltros}, in the
sense that linear expressions for Eqs.~(\ref{F+-}) or
~(\ref{F}), do not correctly describe them. To derive these non-linear
relations within the framework of non-equilibrium
thermodynamics, one may use the internal variables. In this 
alternative picture, one may
consider that the different processes 
 a carrier may go through when it reaches the interface, are
 well represented by a continuous diffusion process in the
 internal space, over an energy barrier
 which matches the discontinuity points at the interface. The carriers
 following this continuous process will be called activated carriers,
 and the coordinate that describes it , the internal coordinate.
 
 By introducing an internal space for each elementary process, whose 
 number
 may be predicted directly from the expression of the entropy 
 production, the
 net rate corresponding to them may be evaluated, allowing one to
 arrive at an explicit expression for Eqs.~(\ref{F+-}) and~(\ref{F}).
 
 It is worth emphasizing that both pictures (internal degrees of
 freedom and macroscopic discontinuity) 
 leads to the same
 entropy production~\cite{Reaccions}, ~\cite{Rose-Mazur}, and it
 is this fact that enables us to use the former to derive explicit
 non-linear constitutive relations, provided the latter one is unable
 to.
 
 \subsection{Thermionic emission relation}
  For the case of unipolar, i.e. those
  with only a single carrier, abrupt semiconductor junctions without
   interface
  states, the surface entropy production may be written as 
  Eq.~(\ref{interno}),
  \begin{equation}
  \sigma_{s}^{s} = - \frac{1}{T} J_{e}^{+} [\phi_{e}]_{-}
  \label{unipolar}
  \end{equation}
  For the sake of definitess, we have taken the electrons as the 
  single carrier and the corresponding magnitudes have been denoted
  by the subindex e. This entropy production corresponds to a single 
  elementary
  process, as can be seen 
  by comparing it with the one corresponding to an unimolecular 
  reaction
  of the form
  \begin{equation}
  e^{-}  \rightleftharpoons e^{+}
  \end{equation}
  provided one identifies $J_{e}^{+}$ with the reaction rate and 
  $[\phi_{e}]_{-}$ with the affinity (see~\cite{Reaccions}, 
  ~\cite{Degroot}).
  Furthermore, this equivalence between chemical reactions and 
  emission 
  processes, allows
  us to directly write the corresponding constitutive relation for 
  the
  latter case by simply identifying the corresponding magnitudes in 
  the
  former case. In this way we obtain, in the
  general case of a degenerate (non-ideal) system,
   ~\cite{Reaccions}, ~\cite{Degroot},
  \begin{equation}
  J_{e}^{+} = k_{B} l \left( 1 - e^{\beta [\phi_{e}]_{-}} \right)
  \label{eqJ}
  \end{equation}
  with
  \begin{equation}
  l= \frac{ e^{\beta \phi_{e}^{-}} }{ k_{B} 
             \int_{\gamma_{1}}^{\gamma_{2}}
   D(\gamma)^{-1} e^{\beta U(\gamma)}
   f(\gamma) \left ( 1 + \frac{\partial \ln f}
   {\partial \ln n} \right) d \gamma}
   \label{lnoideal}
  \end{equation}
  where $k_{B}$ is Boltzman's constant, $\beta = (k_{B} T)^{-1}$,
  $\gamma$ is the internal coordinate, having values in the interval 
  $[ \gamma_{1}, \gamma_{2}]$ of the internal space, $D(\gamma)$ is 
  the diffusion coefficient
  in the internal space, $U(\gamma)$ is the internal energy 
  barrier, $n(\gamma)$
  is the density number of activated electrons at point $\gamma$, and
  $f(\gamma) = f( n(\gamma), T)$ accounts for the
  degenerancy effects, and is defined through
  the state equation
  \begin{equation}
  \phi( \gamma) = k_{B} T \ln f(\gamma) n(\gamma) + U(\gamma)
  \label{noideal}
  \end{equation}
  where $\phi( \gamma)$ is the generalised chemical potential of the
   activated
  electrons. For the case of a non-degenerate system, $f$ is only
  a function of T, which considerably simplifies
  Eqs.~(\ref{lnoideal}) and ~(\ref{noideal}). In Eqs.~(\ref{eqJ})
  and~(\ref{lnoideal}) we have used  $\phi (\gamma_{1}) =
   \phi_{e}^{-}$
  and $\phi (\gamma_{2}) = \phi_{e}^{+}$, for at these points the 
  internal
  space makes contact with the external one. Moreover, in what
  follows, we will also identify 
  $f(n(\gamma_{1}),T) = f_{e}^{-} (n_{e}^{-},T)$, 
  $f(n(\gamma_{2}),T) = f_{e}^{+} (n_{e}^{+},T)$  and 
  $U(\gamma_{1}) = U_{e}^{-}$, $U(\gamma_{2}) = U_{e}^{+}$. In the 
  non-degenerate case $f_{e}^{\pm, nd} = (N_{e}^{\pm})^{-1}$, where
  $N_{e}^{\pm}$ is the effective density of states, which is only a 
  function
  of T. As we have assumed the mass of the carriers to be the same 
  on each
  side of the junction, we will then have that 
  $N_{e}^{-} = N_{e}^{+} \equiv N_{e}$. 
   
  A simplified expression for Eq.~(\ref{lnoideal}) may be obtained by 
  introducing a mobility into the internal space, $\tilde{\mu}$,
   through Einstein's relation
  \begin{equation}
  \frac{D(\gamma)}{\tilde{\mu}(\gamma)}  = 
  n (\gamma) 
  \frac{\partial (\phi(\gamma)- U(\gamma))}{\partial n(\gamma)}  = 
                                          k_{B} T \left ( 1 + 
          \frac{\partial \ln f}{\partial \ln n} \right) 
          \hspace{0.2 cm} \mbox{.}
\end{equation}
 We then obtain
 \begin{equation}
 l = T \frac{ e^{\beta \phi_{e}^{-}}}
                { k_{B} \int_{\gamma_{1}}^{\gamma_{2}}
   \tilde{\mu}(\gamma)^{-1} e^{\beta U(\gamma)} f(\gamma) d \gamma}
  \hspace{0.2 cm} \mbox{.}  \label{l2}
 \end{equation}
 For the purposes of this paper, it is convenient to define a  
 transition coefficient as follows
 \begin{equation}
 \lambda_{-+} \equiv k_{B} \frac{l}{f_{e}^{-} n_{e}^{-}}
 \hspace{0.2 cm} \mbox{.} \end{equation}
 With this definition, and by substituting Eq.~(\ref{l2}) in it, we 
 have
 \begin{equation}
 \lambda_{-+} = T \frac{1}{\int_{\gamma_{1}}^{\gamma_{2}}
   \tilde{\mu}(\gamma)^{-1} e^{\beta (U(\gamma) - U(\gamma_{1}))} 
    f(\gamma) d \gamma}
   \hspace{0.2 cm} \mbox{.} \end{equation}
 Note that for an abrupt junction, $U(\gamma)$ is a given function 
 independent on
 the  density number. Moreover $\tilde{\mu}$ may also be considered 
 independent
 of the density number, even for a degenerate system. Hence, the main
  dependence
 of $\lambda_{-+}$ on the density is given through 
 $f$. As a consequence, for a non-degenerate system, in which $f$ is
  only a 
 function of T, this transition coefficient will be independent 
 on the densities.
 
 Similarly, we may define the inverse transition coefficient, 
 $\lambda_{+-}$,
 as
 \begin{equation}
 \lambda_{+-} \equiv \lambda_{-+} e^{\beta \Delta U}
 \end{equation}
 where $\Delta U = U(\gamma_{2}) - U(\gamma_{1}) = U_{e}^{+} 
 -U_{e}^{-}$
 represents the discontinuity in the electron energy, which for an 
 abrupt 
 junction is independent of the density.
  
  By using this coefficient we may rewrite Eq.~(\ref{eqJ}) as
  \begin{equation}
  J_{e}^{+} = \lambda_{+-} e^{- e \beta \phi_{b}^{+}} \left(1 
                        - e^{\beta [\phi_{e}]_{-}} \right)
  \label{eqJ2}                      
 \end{equation}
 where $e \phi_{b}^{+} = U_{e}^{+} - \phi_{e}^{-}$. It is easily seen 
 that
 Eq.~(\ref{eqJ2}), or alternatively Eq.~(\ref{eqJ}), satisfy the 
 unipolar
 version of the thermodynamic requirements presented in section II,
  Eqs.~(\ref{equil}) and ~(\ref{nonequil2}) ( no symmetry relation is
   needed 
  here). Moreover,
  as predicted in that case, the discontinuity relation only depends 
  on the
  thermodynamic force, in this case $[\phi_{e}]_{-}$, and on 
  $n_{e}^{\pm}$,
  through $\lambda_{+-}$, and $T$. On the other hand, for the 
  non-degenerate
  case, $\lambda_{+-}$, is only a function of $T$, and we denote it 
  by 
  $\lambda_{+-}^{nd}$.
  
  Let us show that Eq.~(\ref{eqJ}), in the non-degenerate case, is 
  precisely
  equivalent to the boundary conditions used in the TE-D 
  theories~\cite{Sze}.
   By simply
  noting that for that case one has
  \begin{equation}
  n_{e}^{\pm} = N_{e} e^{\beta (\phi_{e}^{\pm}-U_{e}^{\pm})}
  \label{nondeg}
  \end{equation}
  we can rewrite Eq.~(\ref{eqJ2}) as 
  \begin{equation}
  e J_{e}^{+} = e \frac{\lambda_{+-}^{nd}}{N_{e}} (n_{m}^{+} -
  n_{e}^{+})
  \label{eqJ3}
  \end{equation}
  where $n_{m}^{+} = N_{e} e^{- e \beta \phi_{b}^{+}}$. If we 
  identify 
  $v_{R} \equiv \frac{\lambda_{+-}^{nd}}{N_{e}}$, we then see that
  Eq.~(\ref{eqJ3}) is precisely the boundary condition used 
  in~\cite{Sze},
  in the context of MS junctions.
  A kinetic calculation for it gives~\cite{Sze} 
  $v_{R} = \frac{A^{*} T^{2}}{e N_{e}}$,
  with $A^{*}$ being Anderson's modified constant. Therefore, we 
  obtain 
  $\lambda_{+-}^{nd} = \frac{A^{*} T^{2}}{e}$, which is ,as mentioned
  above,
  only a function of T.
  
  Moreover, by using Eq.~(\ref{nondeg}) again, we may rewrite 
  Eq.~(\ref{eqJ2})
  as
  \begin{equation}
  e J_{e}^{+} = e \frac{\lambda_{+-}^{nd}}{N_{e}} \left( n_{e}^{-}
                          e^{-\beta \Delta U} - n_{e}^{+} \right)
  \label{eqJ4}                        
  \end{equation}
  which corresponds to the boundary condition proposed 
  in~\cite{llibre} in 
  the context of heterojunctions. 
  It is worth pointing out that Eqs.~(\ref{eqJ3}) and ~(\ref{eqJ4}) 
  reduce, 
  respectively, to Bethe's~\cite{Bethe} and Anderson's~\cite{Anderson}
   expressions, in the limit
  in which diffusion processes are neglected.
  
  \section{Thermionic emission-interface state relations 
                       for abrupt semiconductor junctions}
  
  Our aim in this section is to generalise the results of the 
  previous section, 
  by including
  the effects of interfacial states or by considering
  bipolar systems, i.e. systems with two types of carriers.
  
  \subsection{Unipolar systems with interfacial states}
  
  For this case the 
  surface entropy production can be written as, see 
  Eq.~(\ref{intersi}),
  \begin{eqnarray}
  \sigma_{s}^{s} &=& - \frac{1}{T} J_{e}^{+} X_{e}^{+} 
                 - \frac{1}{T} J_{e}^{-} X_{e}^{-} = \nonumber \\
                 & & - \frac{1}{T} J_{e}^{+} (\phi_{e}^{+} - 
                 \phi_{e}^{s})
                     - \frac{1}{T} (-J_{e}^{-}) (\phi_{e}^{-} - 
                     \phi_{e}^{s}) \hspace{0.2 cm} \mbox{.}
  \end{eqnarray}
  This expression is reminiscent of the entropy production 
  corresponding to
  the set of independent elementary processes concerning the 
  interchange
  of electrons among bulks and surface,
   \begin{eqnarray}
   e^{-} &\rightleftharpoons& e^{s} \hspace{1 cm}  \nonumber  \\
   e^{s} &\rightleftharpoons& e^{+} \hspace{1 cm}  \label{reac} \\   
   e^{-} &\rightleftharpoons& e^{+} \hspace{1 cm} \nonumber
   \end{eqnarray}
   labelled by $\alpha = 1, 2, 3$, 
  with reaction rates $J_{1}= -J_{e}^{-}$, $J_{2}= J_{e}^{+}-
  J_{e}^{-}$,
   $J_{3}= J_{e}^{+}$, and conjugated affinities $A_{1} = 
   X_{e}^{-} $,
   $A_{2} = X_{e}^{+} $, $A_{3} = X_{e}^{-} + X_{e}^{+} = 
   [\phi_{e}]_{-}$.
   
   The equivalence with the chemical reaction case analyzed 
   in~\cite{Reaccions}, 
   allows us to directly formulate  the constitutive relations
   \begin{equation}
   J_{i} = \sum_{\alpha=1}^{3} k_{B} \nu_{i \alpha} l_{\alpha} 
             \left( 1 - e^{\beta A_{\alpha}} \right) \hspace{1 cm} 
             i = 1, 2 , 3
   \label{Ji}
   \end{equation}
   where  $\nu_{i \alpha}$ is the stoichiometric coefficient of the 
   $i$-th
   electron ( we identify $e^{-} \equiv e_{1}$, $e^{s} \equiv e_{2}$
   and $e^{+} \equiv e_{3}$ ) participating in the $\alpha$-th  
   elementary
   reaction, $A_{\alpha}$ are the affinities defined above and
   \begin{equation}
   l_{\alpha}= 
 \frac{e^{\beta \phi_{\alpha} (\gamma_{\alpha_{1}})} }{ k_{B} 
   \int_{\gamma_{\alpha_{1}}}^{\gamma_{\alpha_{2}}}
   D_{\alpha} (\gamma_{\alpha})^{-1} e^{\beta U_{\alpha} 
   (\gamma_{\alpha})}
   f_{\alpha} (\gamma) \left ( 1 + 
   \frac{\partial \ln f_{\alpha}}{\partial \ln n_{\alpha}} \right) 
   d \gamma_{\alpha}}
   \label{lalfa}
  \end{equation}   
   where 
   $\phi_{\alpha} (\gamma_{\alpha_{1}})$
   corresponds to the generalised chemical potential of the 
   {\em reactant} 
   in the  $\alpha-th$ reaction. As before, by introducing a mobility,
   $\tilde{\mu}_{\alpha}$,
   through Einstein's relation we can rewrite Eq.~(\ref{lalfa}) as
   \begin{equation}
   l_{\alpha}= T 
   \frac{ e^{\beta \phi_{\alpha} (\gamma_{\alpha_{1}})} } 
   {\int_{\gamma_{\alpha_{1}}}^{\gamma_{\alpha_{2}}}
   \tilde{\mu}_{\alpha} (\gamma_{\alpha})^{-1} 
   e^{\beta U_{\alpha} (\gamma_{\alpha})}
   f_{\alpha} (\gamma) d \gamma_{\alpha}}
   \hspace{0.2 cm} \mbox{.} \label{lalfa2}
   \end{equation}
   For the elementary processes under consideration, 
   Eq.~(\ref{reac}), we have 
   \begin{equation}
   ( \nu_{i \alpha} ) = \left(
     \begin{array}{rrr}
      -1 &  1 &  0  \\ 
       0 & -1 &  1 \\
       1 &  0 & -1 
      \end{array}
     \right)
     \hspace{0.2 cm} \mbox{.} \label{nus}
  \end{equation}
  By substituting these values into Eq.~(\ref{Ji}), we obtain the 
  following
  constitutive relations,
  \begin{eqnarray}
  J_{e}^{+} & = & k_{B} \left[ l_{2} \left( 1 - e^{\beta X_{e}^{+}} 
  \right) +
 l_{3} \left( 1 - e^{\beta [\phi_{e}]_{-}} \right) \right] 
 \label{Je+} \\
  J_{e}^{-} & = & k_{B} \left[ l_{1} \left( 1 - e^{\beta X_{e}^{-}} 
  \right) +
   l_{3} \left( 1 - e^{\beta [\phi_{e}]_{-}} \right) \right]  
   \hspace{0.2 cm} \mbox{.} \label{Je-}       
  \end{eqnarray}
  Note that the right hand sides of the previous equations are 
  actually functions
  of the thermodynamic forces, $X_{e}^{\pm}$, 
  $[\phi_{e}]_{-}$ (only two of them are independent because
  $[\phi_{e}]_{-} = X_{e}^{+} + X_{e}^{-}$ ), and
  of T and $n_{e}^{s}$. To see this last dependence, we note that 
  although the
   phenomenological
  coefficients, at first sight, may also depend on the bulk densities,
   $n_{e}^{\pm}$, this dependence gives rise to a dependence
   on $n_{e}^{s}$ and $X_{e}^{\pm}$, through the equations of state
   \begin{equation}
   f_{e}^{\pm} n_{e}^{\pm} = e^{ \beta (U_{e}^{\pm} - U_{e}^{s})} 
        e^{ \pm \beta X_{e}^{\pm}} f_{s}^{s} n_{e}^{s}
   \hspace{0.2 cm} \mbox{.}     
   \end{equation}
  Furthermore, by construction, Eqs.~(\ref{Je+}) and~(\ref{Je-})
  satisfy the thermodynamic restrictions given
  through Eqs.~(\ref{equil}), ~(\ref{nonequil1}) 
  and~(\ref{Onsager1}). We therfore conclude that
  they constitute acceptable thermodynamic constitutive relations.
  
  As in section III, a more useful form of these constitutive
  relations is obtained by defining
  the following transition coefficients
  \begin{eqnarray}
  \lambda_{ij} & \equiv & - \sum_{\alpha=1}^{\alpha=3} 
        \frac{ \nu_{i \alpha} l_{\alpha} \nu_{j \alpha}}{z_{i}} 
        \hspace{0.5 cm}
  \mbox{,}       \hspace{0.5 cm} i < j  \label{i<j} \\
  \lambda_{ij} & \equiv & \lambda_{ji} e^{\beta \Delta U_{ij}} 
  \hspace{1.3 cm} 
   \mbox{,}       \hspace{0.5 cm}           i > j
  \label{i>j}
  \end{eqnarray}
  where $z_{i} = f_{i} n_{i}$ are the activities and 
  $\Delta U_{ij} = U_{i} - U_{j}$. We then obtain a master-like 
  equation
  \begin{equation}
  J_{i} = \sum_{j=1}^{j=3} \lambda_{ji} z_{j} - 
          \sum_{j=1}^{j=3} \lambda_{ij} z_{i}  \hspace{1 cm} i=1,2,3 
          \hspace{0.3 cm} \mbox{;}  \hspace{0.3 cm} i \neq j
          \hspace{0.2 cm} \mbox{.}
  \end{equation}
  Note that for an abrupt junction $\Delta U_{ij}$ are given 
  quantities,
  which allow us to write  Eq.~(\ref{i>j}) as a detailed balance -like
  expression, $ \lambda_{ij} z_{i}^{eq} = \lambda_{ji} z_{j}^{eq}$, 
  where
  we have used $ e^{\beta \Delta U_{ij}} =  e^{\beta \Delta 
  U_{ij}^{eq}}=
  \frac{z_{j}^{eq}}{z_{i}^{eq}} $. It should be pointed out that in 
  the 
  degenerate case, as can be seen from Eqs.~(\ref{i<j}) 
  and ~(\ref{i>j}),
  the transition coefficient may depend on the density number, and 
  hence on the
  applied bias. On the other hand, in the non-degenerate case,
  $f_{\alpha}$ becomes independent on the density, and then the 
  transition
  coefficients are only a function of T. In this case, we introduce 
  new
  transition coefficients $\lambda_{ij}^{0} = \lambda_{ij}^{nd} 
  f_{i}^{nd}$,
  where as before we assume $f_{i}^{nd}$ to be independent on i and 
  equal
  to $N_{e}$. Then the constitutive relations transform 
  into~\cite{Reaccions}
  \begin{equation}
  J_{i} = \sum_{j=1}^{j=3} \lambda_{ji}^{0} n_{j} - 
     \sum_{j=1}^{j=3} \lambda_{ij}^{0} n_{i}  \hspace{1 cm} i=1,2,3 
     \hspace{0.3 cm} \mbox{;} \hspace{0.3 cm} i \neq j
  \end{equation}
  with $\lambda_{ij}^{0} n_{i}^{eq} = \lambda_{ji}^{0} n_{j}^{eq}$ and
  $\lambda_{ij}^{0}$ only varying with T.
  
  Keeping these definitions in mind, we may rewrite Eqs.~(\ref{Je+}) 
  and
  ~(\ref{Je-}) as
  \begin{eqnarray}
  J_{e}^{+} = N_{e} \lambda_{31}^{0} e^{ - \beta (U_{e}^{+}-
  \phi_{e}^{-})}
               \left( 1 - e^{\beta [\phi_{e}]_{-}} \right)
           + N_{e} \lambda_{32}^{0} e^{ - \beta (U_{e}^{+}-
           \phi_{e}^{s})}
               \left( 1 - e^{\beta X_{e}^{+}} \right) \label{Je+1} \\
  J_{e}^{-} = N_{e} \lambda_{21}^{0} e^{ - \beta (U_{e}^{s}-
  \phi_{e}^{-})}
               \left( 1 - e^{\beta X_{e}^{-}} \right)
           + N_{e} \lambda_{31}^{0} e^{ - \beta (U_{e}^{+}-
           \phi_{e}^{-})}
               \left( 1 - e^{\beta [\phi_{e}]_{-}} \right) 
               \label{Je-1}             
  \end{eqnarray}
  which constitutes the generalization of Eq.~(\ref{eqJ2}), or 
  \begin{eqnarray}
  J_{e}^{+} =  \lambda_{31}^{0} 
               \left( n_{e}^{-} e^{- \beta (U_{e}^{+} - U_{e}^{s} )} 
               - n_{e}^{+} \right)
           + \lambda_{32}^{0} 
           \left( n_{e}^{s} e^{- \beta (U_{e}^{s} - U_{e}^{-})} - 
               n_{e}^{+} \right) \label{Je+2} \\
  J_{e}^{-} = \lambda_{21}^{0} 
               \left( n_{e}^{-} e^{- \beta (U_{e}^{s} - U_{e}^{+} )} 
               - n_{e}^{s} \right)
           + \lambda_{31}^{0} 
           \left( n_{e}^{-} e^{- \beta (U_{e}^{+} - U_{e}^{-})} - 
               n_{e}^{+} \right) \label{Je-2}            
  \end{eqnarray}
  which also generalizes Eq.~(\ref{eqJ4}). The corresponding
  expressions valid for the degenerate case may be obtained by simply 
  substituting 
  $\lambda_{ij}^{0} N_{e}$ by $\lambda_{ij}$ and $n_{i}$ by $z_{i}$.
  This simple rule then justifies the fact that  in what
  follows only the non-degenerate expressions will be treated 
  explicitly.
  
  Some final remarks are in order. First of all, it is 
  worth noting that Eqs.~(\ref{Je+1}) and~(\ref{Je-1}), or 
  alternatively, 
  Eqs.~(\ref{Je+2}) and~(\ref{Je-2}), 
  hold under non-steady conditions for which $J_{e}^{+} \neq 
  J_{e}^{-}$.
  Moreover,
  they also include all the possible transitions among 
  interface states and bulks and all the situations of interest
  (degenerate and non-degenerate systems). Finally, it is worth
  noting that the number of parameters
  to be specified is relatively small: the transition coefficients
  $\lambda_{21}^{0}$, $\lambda_{31}^{0}$, $\lambda_{32}^{0}$, and
  the energy discontinuities $U_{e}^{+} - U_{e}^{s}$, $U_{e}^{+} - 
  U_{e}^{-}$
  (clearly $U_{e}^{-} - U_{e}^{s}$ is not an independent quantity). 
  This
  makes the modelling of these processes not as difficult as it
  would seem a priori.
  
  \subsection{ Steady-state conditions}
  
  Under steady state conditions Eqs.~(\ref{Je+1}) - ~(\ref{Je-2}) 
  can be
  simplified considerably. In this case one has that $J_{e}^{+} = 
  J_{e}^{-}$,
  see ~\cite{Noltros}, and we then obtain
  \begin{eqnarray}
  J_{e}^{+}          & = & N_{e} \left( \lambda_{31}^{0} + 
    \frac{1}{ (\lambda_{32}^{0})^{-1} + 
    (\lambda_{21}^{0})^{-1} e^{- \beta (U_{e}^{+} - U_{e}^{-})} } 
    \right)
    e^{ - \beta (U_{e}^{+}-\phi_{e}^{-})}
               \left( 1 - e^{\beta [\phi_{e}]_{-}} \right) 
               \label{sJe+1} \\
 e^{\beta X_{e}^{+}} & = & \frac{ 1 + 
 \frac{\lambda_{21}^{0}}{\lambda_{32}^{0}} e^{ \beta (U_{e}^{+} - 
 U_{e}^{s})}}{ 1 + 
 \frac{\lambda_{21}^{0}}{\lambda_{32}^{0}} e^{ \beta (U_{e}^{+} - 
 U_{e}^{s})}
 e^{- \beta [\phi_{e}]_{-}}} \label{sJe-1}
 \end{eqnarray}
 where, as noted in~\cite{Noltros}, the relevent variable 
 under steady state conditions is $[\phi_{e}]_{-}$. Alternatively, we 
 can also
  write
  \begin{eqnarray}
  J_{e}^{+} & = & \left( \lambda_{31}^{0} + 
    \frac{1}{ (\lambda_{32}^{0})^{-1} + 
    (\lambda_{21}^{0})^{-1} e^{- \beta (U_{e}^{+} - U_{e}^{-})} } 
    \right)
   \left(n_{e}^{+} - n_{e}^{-} e^{- \beta (U_{e}^{+} - U_{e}^{-})} 
   \right) 
   \label{sJe+2} \\
  n_{e}^{s} & = & \frac{1}{ \lambda_{21}^{0} + \lambda_{32}^{0}
  e^{- \beta (U_{e}^{+} - U_{e}^{s})} } \left( \lambda_{21}^{0}
  e^{- \beta (U_{e}^{s} - U_{e}^{-})} n_{e}^{-} + 
  \lambda_{32}^{0} n_{e}^{+} \right) \label{sJe-2}
 \hspace{0.2 cm} \mbox{.} \end{eqnarray}
 The corresponding expressions for the degenerate case
 are obtained by using the transformation rule
 formulated previously.
 
 It is worth mentioning that Eqs.~(\ref{sJe+1}) and~(\ref{sJe+2}) 
 have a 
 TE-like form, see Eqs~(\ref{eqJ2}) and ~(\ref{eqJ4}), but they 
 differ in the fact
 that now the prefactor incorporates the effects of the interfacial 
 states.
 Furthermore, we have obtained the equations governing the population
 of those states under non-equilibrium conditions, Eqs.~(\ref{sJe-1}) 
 or ~(\ref{sJe-2}), in their
 general form. From these relations, approximate expressions can be 
 proposed.
 For instance, let us consider the case, very common in the current 
 literature,
 in which it 
 is assumed that the interface states are in permanent equilibrium
 with one of the bulk systems, say system $-$. This situation 
 corresponds
 to the case in which the transitions $ e^{s} \rightarrow  e^{+}$ are
 practically inhibited as compared to the $ e^{s} \rightarrow  e^{-}$
 transitions. In our formulation this means that
 \begin{equation}
 \frac{\lambda_{21}^{0}}{\lambda_{23}^{0}} = 
        \frac{\lambda_{21}^{0}}{\lambda_{32}^{0}} 
        e^{- \beta (U_{e}^{s} - U_{e}^{+})} \gg 1
  \hspace{0.2 cm} \mbox{.}
  \end{equation}
 Using this approximation in Eq.~(\ref{sJe+1}), and considering
 moderate values for $[\phi_{e}]_{-}$, we obtain
 \begin{equation}
 \phi_{e}^{s} \approx \phi_{e}^{-}
 \end{equation}
 which is the condition of permanent equilibrium between the 
 interface
 and the $-$ system, and which ultimately justifies the
 use of the bulk chemical potentials to control
 the filling of the interface states. Moreover, in this case, the 
 prefactor in
 Eq.~(\ref{sJe+2}) reduces to $ \lambda_{31}^{0} + \lambda_{32}^{0}$,
 which is thought to introduce a small modification with respect to 
 the
 commonly used TE-like expressions. It should be noted that for high
 values of $[\phi_{e}]_{-}$ this approximation may not be justified, 
 and
 non-trivial effects coming from the presence of interface states 
 could appear.
 
 \subsection{Bipolar systems}
 In this subsection, we will consider the case of bipolar systems. For
 simplicity's sake we will start by considering the case in which 
 interface states are not present. In this case the surface entropy 
 production
 reads, Eq.~(\ref{interno}),
 \begin{equation}
 \sigma_{s}^{s} = -\frac{1}{T} [\phi_{h}]_{-} J_{h}^{+} 
                  -\frac{1}{T} [\phi_{e}]_{-} J_{e}^{+}
 \end{equation}
 where the subindex h stands for holes and e for electrons. It is 
 easily seen
 that the elementary processes corresponding to this situation are 
 simply
 \begin{eqnarray}
 e^{-} & \rightleftharpoons & e^{+}  \nonumber \\
 h^{-} & \rightleftharpoons & h^{+} 
 \hspace{0.2 cm} \mbox{.} \end{eqnarray}
 Note that these processes are decoupled and hence they can be 
 treated
 separately, proceeding along the lines indicated in section II. The 
 constitutive relations for this case will then read 
 \begin{eqnarray}
  J_{e}^{+} & = & N_{e} \lambda_{+-}^{e 0} 
  e^{- \beta ( U_{e}^{+} - \phi_{e}^{-})} \left(1 
                        - e^{\beta [\phi_{e}]_{-}} \right) = 
                        \label{bipe1}  \\
            &   & \lambda_{+-}^{e 0}
    \left( n_{e}^{-} e^{- \beta (U_{e}^{+} - U_{e}^{-})} - 
    n_{e}^{+} \right) 
    \label{bipe2} \\
  J_{h}^{+} & = & N_{h} \lambda_{+-}^{h 0} 
  e^{- \beta ( U_{h}^{+} - \phi_{h}^{-})} \left(1 
                        - e^{\beta [\phi_{h}]_{-}} \right) = 
                        \label{biph1}  \\
            &   & \lambda_{+-}^{h 0}
    \left( n_{h}^{-} e^{- \beta (U_{h}^{+} - U_{h}^{-})} - 
    n_{h}^{+} \right) 
    \label{biph2}                 
 \end{eqnarray}
 where the meaning of the different magnitudes clearly follows from 
 the
  notation. As before,
 the degenerate case is obtained by applying the corresponding rule. 
 Again
 by construction the corresponding thermodynamic restrictions are 
 satisfied.
 Note that in this case, we have to specify four parameters, the
 transition coefficients $\lambda_{+-}^{e 0}$, $\lambda_{+-}^{h 0}$, 
 and
 the energy discontinuities $U_{e}^{+} - U_{e}^{-}$, $U_{h}^{+} - 
 U_{h}^{-}$.
 For bipolar heterojunctions the two carriers are hardly ever
 considered both together, for usually one of the discontinuities
 is much larger than the other inhibiting the crossing of the 
 corresponding type of carrier~\cite{Milnes}. This fact is 
 obviously recovered from our relations.
 
 The introduction of interfacial states in these systems does not 
 complicate
 the situation very much, because due to the decoupling that exists 
 between the
 two types of carriers, they can be treated separately, giving rise
 to a set of equations like the ones obtained in the previous 
 subsection
 for each of the carriers.
 
 \section{Conclusions}
 In this paper we have presented a fully thermodynamic and general 
 treatment of 
 thermionic emission processes in semiconductor 
 junctions.
 All the situations of interest, namely, for unipolar and bipolar
 systems, which can be degenerate or not, with or without interface 
 states and
 under steady and non-steady conditions, have been analysed.
 
 The study has been carried out within the framework of 
 non-equilibrium
  thermodynamics of systems with interfaces, incorporating the 
  formalism
  of the internal degrees of freedom in order to obtain  explicit 
  expressions
  for the TE-I relations. This procedure has the advantage of 
  allowing
  us  a complete treatment of semiconductor junctions (bulks,
  interface and discontinuity relations) in a common framework which
  ensures by construction the thermodynamic consistency and 
  completeness
  of the relations derived.
  
  Of special interest has been the derivation of general TE-IS 
  relations
  to describe non-steady situations in the presence of interface
  states. From these expressions we have then derived the 
  corresponding 
  relations valid at the steady state, which have been shown to 
  incorporate, in
   the general
  case, non-trivial contributions that come from the interface 
  states, which
  might not be negligible in some circumstances. These interface 
  state
  contributions are under investigation for some junctions of 
  interest.
  
  Finally, the implementation of the TE-IS expressions in the 
  drift-diffusion
  model presented in~\cite{Noltros}, will provide one of the most
  general TE-IS-D models that has been proposed for abrupt 
  semiconductor 
  junctions, proving therefore the power of the application of the 
  non-equilibrium
  thermodynamic methods to semiconductor systems. 

\acknowledgements
This work has been supported by DGICYT of the Spanish Government"
under grant PB92-0859 and the European Union Human Capital and
Mobility Programme, contract ERB-CHR-XCT93-0413.
One of us (G.G.) wishes
to thank CIRIT of Generalitat de Catalunya for financial support.

\end{document}